\documentclass[aps,prd,tightenlines,preprint,nofootinbib,a4paper]{revtex4}
\usepackage{epsfig}

\begin{document}

\title{Signatures of the neutrino mass hierarchy in supernova neutrinos}

\author{S.~H.~Chiu$^{1}$\footnote{$schiu@mail.cgu.edu.tw$}}
\author{Chu-Ching Huang$^{2}$\footnote{$cchuang@mail.cgu.edu.tw$}}
\author{Kwang-Chang Lai$^{1,3}$\footnote{$kcl@mail.cgu.edu.tw$}}

\affiliation{%
$^{1}$Physics and $^{2}$Mathematics Groups, CGE, Chang Gung University, Kwei-Shan 333, Taiwan \\
 \\ 
$^{3}$Leung Center for Cosmology and Particle Astrophysics (LeCosPA),
National Taiwan University, Taipei, 106, Taiwan}


\begin{abstract}

The undetermined neutrino mass hierarchy may leave an observable imprint on the 
neutrino fluxes from a core-collapse supernova (SN).
The interpretation of the observables, however, is subject to the uncertain SN models and 
the flavor conversion mechanism of neutrinos in a SN. 
We attempt to propose a qualitative interpretation of the expected neutrino events
at terrestrial detectors, focusing on the accretion phase of the neutrino burst.
The flavor conversions due to neutrino self-interaction, the MSW effect,
and the Earth regeneration effect are incorporated in the calculation.
It leads to several distinct scenarios that are identified by the neutrino mass hierarchies
and the collective flavor transitions. 
Consequences resulting from the variation of incident angles and SN models are also discussed.

\end{abstract}


\maketitle

\pagenumbering{arabic}



\section{Introduction}

Our knowledge of the neutrino ($\nu$) flavor transition  
in a core-collapse supernova (SN) has
encountered a paradigm shift in the recent years. 
It was pointed out (for an incomplete list, see, e.g., Refs. \cite{Pantaleone:1992eq,Duan:2010bg}) that  
in the deep region of the core where neutrino densities are large,
neutrino self-interaction could result in significant flavor conversion 
that is totally distinct in nature from the 
well-known Mikheyev-Smirnov-Wolfenstein (MSW) effect \cite{Wolfenstein:1977ue+Mikheev:1986gs},  
which instead arises from the interaction of neutrinos with the ordinary stellar medium.
The intense study 
has suggested that coherent $\nu-\nu$ 
forward scatterings may lead to collective pair conversion $\nu_{e}\bar{\nu}_{e} \leftrightarrow \nu_{x}\bar{\nu}_{x}$
($x=\mu,\tau$) over the entire energy range even with extremely small mixing angles.
It was also pointed out that 
in a typical supernova, this type of flavor conversion would take place near $r \sim 10^{3}$ km,
in contrast to that for the MSW effect at $r \sim 10^{4}-10^{5}$ km. 
In addition, the development of the collective effects
depends crucially on the primary $\nu$ spectra \cite{Fogli:2007bk,Dasgupta:2009mg},  
as well as on the $\nu$ mass hierarchy. 
One would thus expect non-trivial modifications to the original SN $\nu$ fluxes as they propagate outward.
In general, the self-induced flavor conversion does not alter both $\nu$ and $\bar{\nu}$ spectra
under the normal hierarchy (NH). However, it leads to the nearly complete spectra exchange 
$\bar{\nu}_{e} \leftrightarrow \bar{\nu}_{x}$ and a partial swap of the spectra, $\nu_{e} \leftrightarrow \nu_{x}$, 
at a critical energy under the inverted hierarchy (IH).  
The rich physical content, in a sense, leads to further 
theoretical uncertainties which complicate the interpretation of the
observed SN $\nu$ bursts and the unknown properties of $\nu$
that may be revealed by the observation.

With the unique production and detection processes, the neutrino burst from a SN has
long been considered as one of the promising tools for the study of neutrino parameters and the SN mechanism.
A core-collapse SN emits all three flavors of neutrino with a 
characteristic energy range and a time scale that are totally distinct
from those of the neutrinos emitted from the Sun, the atmosphere, and terrestrial sources.
It has been suggested (see, i.e., Ref.\cite{Raffelt:2010zza}) 
that analyzing SN neutrino bursts may provide   
hints to the unknown elements of the neutrino mass and mixing matrices.
With the recent pin-down of
the last mixing angle $\theta_{13}$ \cite{An:2012eh,Ahn:2012nd}, 
the determination of the neutrino mass hierarchy may seem the next reachable goal. 
Recent study based on multi-angle analysis of SN neutrinos \cite{Chakraborty:2011nf,Sarikas:2011am}, however, 
pointed out that the seemingly dominating collective effects
may be suppressed by the dense matter during the accretion phase following the core bounce \cite{EstebanPretel:2008ni}. 
This time-dependent variation of the neutrino survival probability during the early phase 
(post-bounce times $t_{pb} \lesssim 0.6$s)
would give rise to a new interpretation of the observed neutrino flux and hints to 
the $\nu$ properties, such as the mass hierarchy.

Focused on this issue, we analyze the expected SN $\nu$ events during the early accretion phase
at Earth-bound detectors through two channels of neutrino interactions: 
$\bar{\nu}_{e}+p$ in the water Cherenkov detector (WC) or the scintillation detector (SC), and
$\nu_{e}+ Ar$ in the liquid argon chamber (Ar). 
Given the uncertainties among the SN models,  we propose observables that
may be useful in the determination of the $\nu$ mass hierarchy. 
The outcomes with different incident angles at the detectors are also compared.

This work is organized as follows.
In Section II, we outline the recent progress on the measured neutrino mixing angles,
the squared mass differences, and the general features of the neutrino fluxes emitted by a core-collapse
supernova.  The known parameters will be adopted as the input for the calculation.
Section III is devoted to investigating the modification of the primary neutrino fluxes by 
the collective effect, the MSW effect, and the
Earth regeneration effect.  For illustrative purposes, the calculation is based on a two-layer model
for the Earth matter.  In Section IV, we propose physical observables that may provide hints
to identifying the neutrino mass hierarchy and a working scenario for flavor 
transition due to 
self-interaction.  Expected trends of the event rates at the WC, SC, and Ar detectors
are estimated and discussed.  Results arising from varied incident angles at the detectors
under different SN models are also discussed. 
We then summarize this work in Section V.


\section{Neutrino properties and SN parameters}

The three mixing angles in the $\nu$ mixing matrix have been determined with convincing
precision \cite{data}:  
$\sin^{2}2\theta_{12}\simeq 0.857$, $\sin^{2}2\theta_{23}\geq 0.95$, and
$\sin^{2}2\theta_{13}\simeq 0.098$. The mass-squared differences also have 
been measured: $\delta m^{2}_{21}\simeq 7.6 \times 10^{-5}$ eV$^{2}$,
$|\delta m^{2}_{31}|\simeq 2.4 \times 10^{-3}$ eV$^{2}$.
The absolute neutrino mass, the CP phase in the neutrino sector, and the neutrino mass hierarchy are yet to
be determined.

The primary SN neutrino energy spectrum is typically not pure thermal, and can be modeled as a
pinched Fermi-Dirac distribution \cite{Totani:1997vj}. 
For each neutrino flavor $\nu_{\alpha}$ ($\alpha=e,\mu,\tau$),  
\begin{equation}
F^{0}_{\alpha}=\frac{\phi_{\alpha}}{T^{4}_{\alpha}g(\eta_{\alpha})}\frac{E^{2}_{\alpha}}
{\exp[(E_{\nu}/T_{\alpha})-\eta_{\alpha}]+1},
\end{equation} 
where $\phi_{\alpha}=L_{\alpha}/\langle E_{\nu}\rangle$ is the number flux of $\nu_{\alpha}$,   
with $L_{\alpha}$ the energy luminosity and $\langle E_{\nu} \rangle$ the mean neutrino energy.
$T_{\alpha}$ is the effective temperature
of $\nu_{\alpha}$ inside the respective neutrinosphere, $g(\eta_{\alpha})$ is the normalization factor, and
$\eta_{\alpha}$ is the pinching parameter.  Note that another widely adopted parametrization is given by, see, e.g.,
Ref.\cite{Keil:2002in}. 

In general, equipartition of the luminosity among the primary neutrino flavors is expected 
in typical SN simulations, $L^{0}_{\nu_{e}} \approx \bar{L}^{0}_{\nu_{e}}\approx L^{0}_{\nu_{x}}$,
which is assumed in our calculation for the neutrino fluxes during the accretion phase.
Note, however, that variations from this nearly degenerate scenario are also suggested in the literature, 
e.g., $L^{0}_{\nu_{e}}/L^{0}_{\nu_{x}} \sim 0.5-2$, 
$L^{0}_{\nu_{e}}=L^{0}_{\bar{\nu}_{e}}$, $L^{0}_{\nu_{x}}=L^{0}_{\bar{\nu}_{e}}$.
As input parameters, the effective temperatures are fixed in this work: $T_{\nu_{e}}=3$ MeV, $T_{\bar{\nu}_{e}}=5$ MeV,
and $T_{\nu_{x}}=T_{\bar{\nu}_{e}}=7$ MeV. In addition, the pinching parameters are taken to be
$\eta_{\nu_{e}}=3$, $\eta_{\bar{\nu}_{e}}=2$, and $\eta_{\nu_{x}}=\eta_{\bar{\nu}_{x}}=1$.

\begin{table*}[ttt]
 \centering
	\begin{center}
 \begin{tabular}{cccc}  
 
 \hline
 
   (a) &  $\bar{P}_{\nu} \simeq 1$  &  $P_{\nu} \simeq 1$  &  normal \\ 
   (b) & $\bar{P}_{\nu} \simeq \bar{P}_{\nu}(t)$ & $P_{\nu} \simeq P_{s}$  & inverted \\
   (c) & $\bar{P}_{\nu} \simeq 0$ & $P_{\nu} \simeq P_{s}$ & inverted \\
   (d) & $\bar{P}_{\nu} \simeq 1$ & $P_{\nu} \simeq P_{s}$  &  inverted \\
      
   \hline
  \end{tabular}
    \caption{Possible scenarios of the flavor conversion due to the collective effect and
    the mass hierarchies.  
    Note that the time-dependent $\bar{P}_{\nu}(t)$ of case (b) 
    is given in Fig.1, and the step-like function $P_{s}$ is given in Section III.}
  \end{center}
 \end{table*}

\section{Neutrino flavor conversion in SN and Earth}

Despite the variation of SN models, the consensus seems to suggest
a partial to complete modification (depending on the mass hierarchy) 
of the neutrino primary spectra by the collective effect,
which occurs near $r \sim 10^{3}$ km as the neutrinos propagate outwards.
The MSW effect then takes place at $r\sim 10^{4}-10^{5}$ km.  The two effects are considered to
be independent because of the wide separation in space.  The neutrino fluxes get further
modification by the Earth matter before their detection. As far as the $\nu$ properties are concerned,
the advantage of analyzing the $\nu$ bursts at  
an early stage is conspicuous. 
The complicated shock wave does not play a role
in the flavor conversion during this early accretion phase \cite{Serpico:2011ir}. 
The phenomenon, however, becomes more complicated during the later cooling phase when the shock wave
is taken into consideration.
Moreover, while the $\bar{\nu}_{e}$ and $\bar{\nu}_{x}$ spectra are expected to be well separated in the early phase,
they tend to become indistinguishable later
during the cooling phase, and the flavor conversion effect for the $\bar{\nu}_{e}$ channel
becomes difficult to observe.

\begin{figure}[ttt]
\caption{The approximated time-dependent survival probability $\bar{P}_{\nu\nu}(t)$ for case (b).} 
\centerline{\epsfig{file=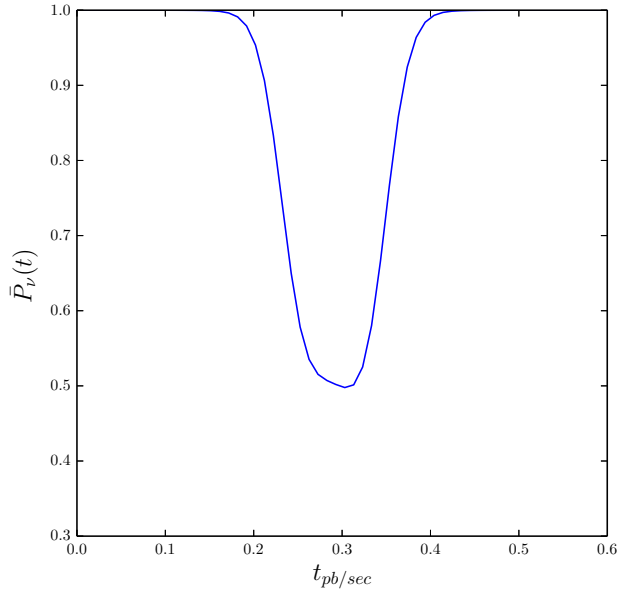,width=9 cm}}
\end{figure}

Recent simulations suggest quite diverse features for the details of the collective effects.
In Table I, we summarize the possible scenarios resulting from the collective effects under both the normal (NH)
and the inverted (IH) hierarchies, 
where $P_{\nu}$ and $\bar{P}_{\nu}$ represent
the survival probabilities of the original $\nu_{e}$ and $\bar{\nu}_{e}$ fluxes, respectively,
after the adjustment of the $\nu\nu$ self-induced collective effects.  There are four different scenarios.
Under NH, both $\nu_{e}$ and $\bar{\nu}_{e}$ spectra remain 
unaltered, i.e., $P_{\nu} \simeq 1$ and $\bar{P}_{\nu}\simeq 1$. This corresponds to case (a).
The IH leads to cases (b), (c), and (d), in which the conversion probability for the $\nu_{e}$ flux can be approximated
by the same step-like function of energy: $P_{s} \simeq 1$ for $E<E_{c}$, and $P_{s} \simeq 0$ for $E>E_{c}$,
where $E_{c} \simeq 8$ MeV is the critical energy \cite{Fogli:2008pt}.  
The three later cases differ in the expected $\bar{\nu}_{e}$ flux.
Case (b) represents partial matter suppression of the collective effect, 
and the survival probability is time dependent: 
$\bar{P}_{\nu}=\bar{P}_{\nu}(t)$ \cite{Chakraborty:2011gd,Chakraborty:2011nf,Sarikas:2011am}. 
Case (c) represents a total swap of the spectra $\bar{\nu}_{e} \leftrightarrow \bar{\nu}_{x}$: 
$\bar{P}_{\nu}\simeq 0$. In this case, the self-induced
effect dominates over the MSW effect, see, e.g., Ref. \cite{Duan:2010bg} .  
We add in our analysis the case (d), which indicates a complete matter suppression of the collective effect: 
$\bar{P}_{\nu}\simeq 1$. This corresponds to the traditional treatment of the SN $\nu$ flux 
based on the pure MSW effect, see, e.g., Ref. \cite{Dighe:1999bi}. 
Note that in our calculation for scenario (b), we adopt an approximate probability function $\bar{P}_{\nu}(t)$ 
(as shown in Fig.1), which is fitted by the results of Ref. \cite{Chakraborty:2011gd}. 


Formulating the neutrino fluxes at different stages is straightforward.
The primary neutrino fluxes are denoted by $F_{e}^{0}$, $F_{x}^{0}$, $\bar{F}_{e}^{0}$, and $\bar{F}_{x}^{0}$.
The first modification to the $\nu$ fluxes in the SN comes from the collective effect, and the fluxes become
\begin{equation}
F^{c}_{e}=F^{0}_{e}+(1-P_{\nu})(F^{0}_{x}-F^{0}_{e}),
\end{equation}
\begin{equation}
F^{c}_{x}=(1-P_{\nu})F^{0}_{e}+(1+P_{\nu})F^{0}_{x},
\end{equation}
\begin{equation}
\bar{F}^{c}_{e}=\bar{F}_{e}^{0}+(1-\bar{P}_{\nu})(\bar{F}^{0}_{x}-\bar{F}^{0}_{e}),
\end{equation}
\begin{equation}\label{eq:fac}
\bar{F}^{c}_{x}=(1-\bar{P}_{\nu})\bar{F}^{0}_{e}+(1+\bar{P}_{\nu})\bar{F}^{0}_{x}.
\end{equation}
Note that $P_{\nu}$ and $\bar{P}_{\nu}$ are given by Table I for varied scenarios and mass hierarchies.

As the neutrinos continue to propagate outwards, they encounter a further modification 
by the MSW effect in the SN.
If one denotes the survival probability for $\nu_{e}$ ($\bar{\nu}_{e}$) after the MSW effect as $P_{m}$ ($\bar{P}_{m}$),
then the fluxes of $\nu_{e}$ and $\bar{\nu}_{e}$ arriving at Earth can be written as 
\begin{equation}\label{eq:fen}
F_{e}=P_{m}F^{c}_{e}+(1-P_{m})F^{c}_{x},
\end{equation}
\begin{equation}
\bar{F}_{e}=\bar{P}_{m}\bar{F}^{c}_{e}+(1-\bar{P}_{m})\bar{F}^{c}_{x},   
\end{equation}
with
\begin{equation}
P_{m}=|U_{e1}|^{2}P_{H}P_{L}+|U_{e2}|^{2}(P_{H}-P_{H}P_{L})+|U_{e3}|^{2}(1-P_{H}),
\end{equation}
\begin{equation}
\bar{P}_{m}=|U_{e1}|^{2}(1-\bar{P}_{L})+|U_{e2}|^{2}\bar{P}_{L},
\end{equation}
for the normal hierarchy, and
\begin{equation}
P_{m}=|U_{e1}|^{2}P_{L}+|U_{e2}|^{2}(1-P_{L}),
\end{equation}
\begin{equation}
\bar{P}_{m}=|U_{e1}|^{2}\bar{P}_{H}(1-\bar{P}_{L})+|U_{e2}|^{2}\bar{P}_{H}\bar{P}_{L}+|U_{e3}|^{2}(1-\bar{P}_{H}),
\end{equation}
for the inverted hierarchy. Here, $P_{H}$ and $P_{L}$ are the crossing probabilities for the neutrino eigenstates at
higher and lower resonances, respectively, and the quantity with a bar represents
that for $\bar{\nu}$.


After propagating through the Earth matter, 
the expected $\nu$ fluxes at the detectors then follow.  
It should be pointed out that, as compared to the analysis of Ref. \cite{Dighe:1999bi},   
we further consider in this work the collective effect which occurs prior to the MSW effect in the SN.
Thus, in deriving the related formulation we may simply replace $F_{e}^{0}$, $F_{\bar{e}}^{0}$,
$F_{x}^{0}$, and $F_{\bar{x}}^{0}$
in Ref. \cite{Dighe:1999bi} by $F_{e}^{c}$, $\bar{F}_{e}^{c}$, $F_{x}^{c}$, and
$\bar{F}_{x}^{c}$ of this work, respectively.
We list the results here and outline the derivations in Appendix A.
For the $\nu_{e}$ flux, we have
\begin{equation}
F^{D}_{e} \simeq F^{0}_{e}[(1-|U_{e3}|^{2})-P_{\nu}(1-2|U_{e3}|^{2})]+
                 F^{0}_{x}[1+P_{\nu}(1-2|U_{e3}|^{2})],
\end{equation}
under the normal hierarchy, and
\begin{equation}
F^{D}_{e} \simeq F^{0}_{e}[(1-P_{2e})-P_{\nu}(1-2P_{2e})]+
                 F^{0}_{x}[1+P_{\nu}(1-2P_{2e})],
\end{equation}
under the inverted hierarchy. 
As for the $\bar{\nu}_{e}$ flux, we get
\begin{equation}
\bar{F}^{D}_{e} \simeq \bar{F}^{0}_{e}[(1-\bar{P}_{1e})-\bar{P}_{\nu}(1-2\bar{P}_{1e})]+
   \bar{F}^{0}_{x}[1+\bar{P}_{\nu}(1-2\bar{P}_{1e})],
\end{equation}
for the normal hierarchy, and 
   \begin{equation}
\bar{F}^{D}_{e} \simeq \bar{F}^{0}_{e}(1-\bar{P}_{\nu})+\bar{F}^{0}_{x}(1+\bar{P}_{\nu}),
\end{equation}
for the inverted hierarchy. 
Here $P_{ie}$ ($\bar{P}_{ie}$) is the probability that a mass eigenstate 
$\nu_{i} (\bar{\nu}_{i}$) is observed as a $\nu_{e}$ ($\bar{\nu}_{e}$)
at the detector. 
Note that we have used the approximation: $P_{3e}-|U_{e3}|^{2} \leq 10^{-3}$ \cite{Dighe:1999bi} 
in writing the detected $\nu$ flux.
Furthermore, we have set vanishing crossing probabilities in writing the above expressions: 
$P_{H}\simeq \bar{P}_{H} \simeq P_{L} \simeq \bar{P}_{L} \simeq 0$.
In contrast to the matter density profile with $\rho \sim r^{-3}$ in the traditional treatment,
it has been pointed out \cite{Chiu:2008sa} that a local deviation of the uncertain density profile from 
$\rho \sim r^{-3}$ may lead to significant change of the MSW crossing probabilities for a certain range
of $\theta_{13}$.  This factor has been taken into consideration in our analysis.
With the recent determination of the relatively large $\theta_{13}$,
we conclude that the vanishing crossing probabilities 
can be adopted
even if the local density profiles near the resonance become as steep as $\rho \sim r^{-8}$.

The probability $P_{2e}$ is usually written as $P_{2e}=\sin^{2}\theta_{12}+f_{reg}$ , 
with $f_{reg}$ the regeneration factor due to the Earth matter effect \cite{deHolanda:2004fd}: 
\begin{equation}
f_{reg}=\frac{2E\sin^{2}2\theta}{\delta m^{2}_{21}}\sin\Phi_{0}\sum_{i=0}^{n-1}\Delta V_{i}\sin\Phi_{i},
\end{equation}
where $n$ is the number of layers, $\Delta V_{i} \equiv V_{i+1}-V_{i}$ 
is the potential difference between adjacent layers of matter,  
and $\Phi_{i}$ is the phase acquired along the trajectories.
For illustrative purposes, we adopt a simple two-layer model \cite{Freund:1999vc} for the Earth matter, 
$\rho_{E}=5.0 \mbox{g/cm}^{3}$ for $R_{\oplus}/2<r<R_{\oplus}$ (mantle) and
$\rho_{E}=12.0 \mbox{g/cm}^{3}$ for $r<R_{\oplus}/2$ (core).  Note that this two-layer analysis can be easily 
generalized to the analysis of a multi-layer model.  The regeneration factor for two layers takes the form,
\begin{equation}
f_{reg}=\sin^{2}2\theta \sin\Phi_{0}[\epsilon_{m}\sin\Phi_{0}+(\epsilon_{c}-\epsilon_{m}) \sin\Phi_{1}],
\end{equation}
with $\epsilon_{m}=2EV_{m}/\delta m^{2}_{21}$ for the mantle and $\epsilon_{c}=2EV_{c}/\delta m^{2}_{21}$
for the core, and 
\begin{equation}
\Phi_{0}=\frac{\delta m^{2}L}{4E}\sqrt{(\cos2\theta-\epsilon_{m})^{2}+\sin^{2}2\theta},
\end{equation}
\begin{equation}
\Phi_{1}=\frac{\delta m^{2}L_{1}}{4E}\sqrt{(\cos2\theta-\epsilon_{c})^{2}+\sin^{2}2\theta}.
\end{equation}
Here $L$ is the total path length inside Earth and $L_{1}$ is that inside the core.

\section{Physical observables in the detectors}

During the accretion phase,
the detailed time evolution of the neutrino fluxes 
is, in general, model dependent. 
In this section we first adopt the ${\nu}$ and $\bar{\nu}$ fluxes 
during the early phase of a $10.8M_{\odot}$ progenitor
SN model, such as in Ref.\cite{Chakraborty:2011nf} where 
the flux evolution during the accretion phase was given up to 0.6s after the core bounce.
We shall later discuss the possible impacts
due to different choices of SN models. 

For our purpose of analyzing the time structure of the energy-integrated fluxes,   
\begin{eqnarray} 
\bar{N}(t) & \sim & \int \bar{F}^{D}_{e}(E,t)\cdot \bar{\sigma}_{\nu}(E)\cdot \bar{\varepsilon}(E) dE, \nonumber \\ 
N(t) & \sim & \int F^{D}_{e}(E,t)\cdot \sigma_{\nu}(E)\cdot \varepsilon(E) dE, 
\end{eqnarray}
we approximate the time-dependent number flux in Eq.(1)  
as piecewise functions (with arbitrary scales) based on Fig. 1 of Ref.\cite{Chakraborty:2011nf}.
Note that since only the flux trend around the accretion phase is relevant to our study,
we omit the early flux details prior to $t_{pb} \sim 0.06$s.
The fitted curves are
\begin{equation}
\phi_{\nu_{e}} = \left\{ \begin{array}{ll}
2.52 \exp[\frac{-(t-0.13)^{2}}{0.065}], & (0 < t \leq 0.25s)  \\
               1.86+\exp[42.95t-13.30], &  (0.25s < t \leq 0.30s) \\ 
                 0.35+\exp[-16.65t+5.73], &  (0.30s <t \leq 0.60s)
                  \end{array}
                  \right.
\end{equation}
\begin{equation}
\phi_{\bar{\nu}_{e}}=\left\{\begin{array}{ll}
 1.99 \exp[\frac{-(t-0.15)^{2}}{0.051}], &  (0< t \leq 0.25s) \nonumber \\
               1.54+\exp[35.74t-11.42], &  (0.25s < t \leq 0.30s) \nonumber \\ 
                0.35+\exp[-16.24t+5.45], &  (0.30s < t \leq 0.60s)
                \end{array}
                \right.
\end{equation}
\begin{eqnarray}
\phi_{\nu_{x}}=\phi_{\bar{\nu}_{x}} = 0.13+\exp[-3.46t+0.037], & (0 < t \leq 0.60s).
\end{eqnarray}
As a comparison for the qualitative features, 
we show in Fig.2 both the fitted curves, Eqs.(21-23), and the data points taken 
from the referential curves. 

We assume a unity
efficiency function, $\bar{\varepsilon}(E)\sim \varepsilon(E) \sim 1$, 
and that the limited energy resolution of the detectors is capable of
observing the qualitative time evolution of the $\nu$ fluxes. The cross-section functions, 
$\bar{\sigma}_{\nu}(E)$ and $\sigma_{\nu}(E)$,
will be presented later for each of the reaction channels.
One may then check the qualitative behavior of the time structures that are the direct
consequences of the mass hierarchies and distinct scenarios of the self-induced effects, 
as will be discussed in the following subsections.


\begin{figure}[ttt]
\caption{The curves representing Eqs.(21-23) for a $10.8 M_{\odot}$ model are compared with the data points taken from the 
referential curves.} 
\centerline{\epsfig{file=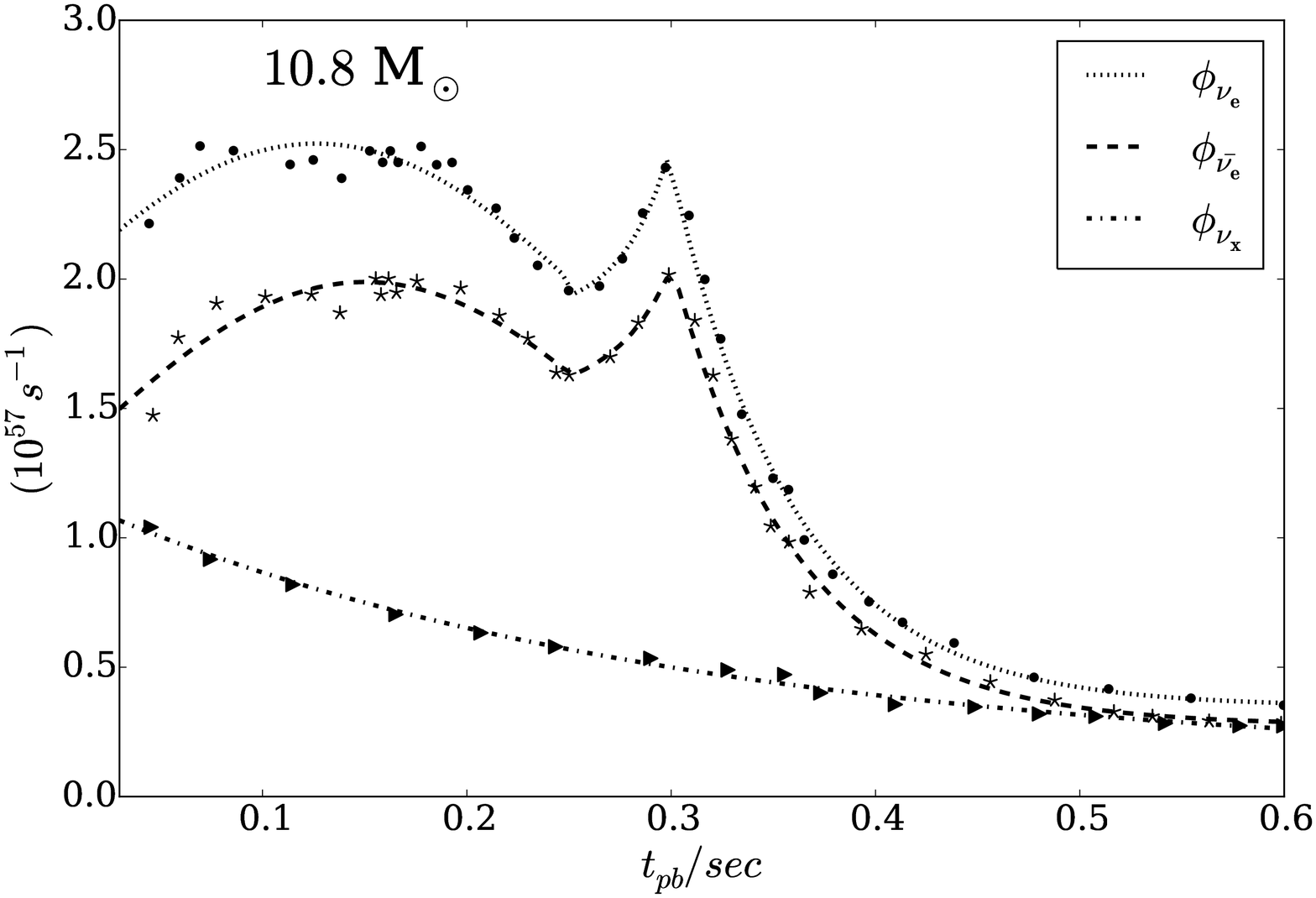,width=11 cm}}
\end{figure}

\begin{figure}[ttt]
\caption{An estimation of $\bar{N}(t)$ (with arbitrary normalization scale) 
at WC or SC during the accretion phase for a $10.8 M_{\odot}$ model. 
The four scenarios, (a), (b), (c), and (d) are resulted
from different mass hierarchies and time structures of the neutrino fluxes, as listed in Table I.
We have used $9\pi/10$ as the zenith
angle at the detector.  Note that a hypothetical scenario labeled by NH$_{0}$ is added as a reference.
Note also that the outcomes for the normal hierarchy are represented by the solid lines,
while that for the inverted hierarchy are represented by the dashed lines.} 
\centerline{\epsfig{file=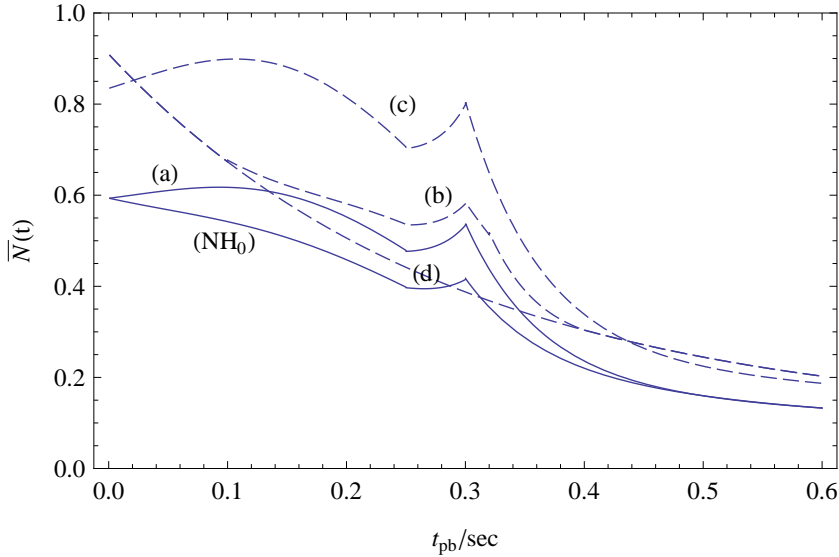,width=11 cm}}
\end{figure}

\subsection{$\bar{\nu}_{e}$ events at WC or SC}

The inverse $\beta$ decay, $\bar{\nu}_{e}+p \rightarrow n+e^{+}$, dominates over other reaction channels
in the water Cherenkov detectors
and the scintillation detectors. We adopt the cross section \cite{Cadonati:2000kq}:
$\sigma(\bar{\nu}_{e}p) \simeq 9.5\times 10^{-44}(E-1.29)^{2} \mbox{cm}^{2}$, and assume that
events originating from this channel can be properly identified.
The time evolution curves of $\bar{N}(t)$ under the four possible scenarios in Table I 
and a hypothetical scenario (NH$_{0}$) are shown in Fig.3, 
and summarized as follows.  

(i) A unique, monotonically decreasing $\bar{N}(t)$, which represents case (d) in Table I,
is easily singled out. Observation of this result not only suggests inverted hierarchy,
but also a near total suppression of the collective effect: $\bar{P}_{\nu}\simeq 1$.
This result favors the scenario that MSW effect $\gg$ collective effect, which
is just the traditional treatment based on the pure MSW effect without the collective effect.

(ii) An early decreasing $\bar{N}(t)$ could also arise from the time-varying survival 
probability $\bar{P}_{\nu}(t)$ as in case (b),
which is due to partial suppression of the collective effect under the inverted hierarchy.  
However, a significant peak at
$t_{pb} \sim 0.3 s$ distinguishes this scenario from that of case (d).

(iii)A slight increase of $\bar{N}(t)$ prior to $t_{pb}\sim 0.1 s$ could represent two different scenarios:
normal hierarchy with $\bar{P}_{\nu} \simeq 1$ (complete suppression of the collective effect, case (a)), or
inverted hierarchy with $\bar{P}_{\nu} \simeq 0$ (total spectrum swap, case (c)).
Although in general the event rates could differ by roughly 50\% according to a quick estimation, 
a definite separation between these two scenarios
may not be easy if one practically considers the model uncertainties and limitations of the experimental resolution.  
We therefore will not stress the significance of this case here.
However, further hints may be available from the observation of the $\nu_{e}$ events.

(iv) Since the overall knowledge of the oscillation effects for SN neutrinos is still insufficient, 
we add in Fig.3 a hypothetical case NH$_{0}\equiv$ (NH,$\bar{P}_{\nu} \simeq 0$), even though current models in the literature
do not favor this scenario.  Note that, as indicated by Table I, the three scenarios (NH,$\bar{P}_{\nu} \simeq 1$), 
(IH,$\bar{P}_{\nu} \simeq 0$), and (IH,$\bar{P}_{\nu} \simeq 1$) are represented by cases (a), (c), and (d), respectively.

The general properties of Fig.3 can be understood as follows:
\begin{itemize}
\item For cases (a), (b), and (c), the peak at $t_{pb} \sim 0.3$s signals the
onset of explosion.  The sharp drop of $\bar{N}(t)$ and the luminosity afterwards is due to
the sudden flip of matter velocities from infall to expansion when
the explosion shock passes through a co-moving frame of reference where the observables
are measured \cite{Chakraborty:2011gd}.
\item With the vanishing $\bar{P}_{H}$ and $\bar{P}_{L}$, the expected $\bar{\nu}_{e}$ flux
at the detector simply inherits the shape of the $\bar{\nu}_{x}$ flux after the modification of 
the collective effect but
before that of the MSW effect: $\bar{F}^{c}_{x}$.  Eq.~(\ref{eq:fac}) suggests that $\bar{F}^{c}_{x}$ is a combination of
$\bar{F}^{0}_{e}$ and $\bar{F}^{0}_{x}$, and the weight of each component is determined by
$1-\bar{P}_{\nu}$ and $1+\bar{P}_{\nu}$, respectively.  This leads to curve (d) in Fig.3 when
$\bar{P}_{\nu}\simeq 1$, which corresponds to the situation when the collective effect is turned off
and only the pure MSW effect is in action. In this case, the expected $\bar{F}^{0}_{e}$
simply follows the monotonically decreasing shape of $\bar{F}^{0}_{x}$ that is common in different SN models.
\item As $\bar{P}_{\nu}$ decreases from $\bar{P}_{\nu}=1$, it represents the situation
when the flavor transition due to the collective effect becomes more important, 
and the fraction of $\bar{F}^{0}_{e}$ becomes
larger.  The direct consequence is that the shape of $\bar{F}^{0}_{e}$ becomes more prominent 
and thus the peak, which is the characteristic feature of $\bar{F}^{0}_{e}$, begins to emerge as in cases (b) and (c)
under the inverted hierarchy.
\item The shape of case (a) under the normal hierarchy can be understood by the same reasoning.
With $\bar{P}_{L} \sim 0$, Eqs.(7) and (9) suggest that 
$\bar{F}_{e} \simeq |U_{e1}|^{2}\bar{F}^{c}_{e}+(1-|U_{e1}|^{2})\bar{F}^{c}_{x}$, i.e.,
$\bar{F}_{e}$ is a superposition of $\bar{F}^{c}_{e}$ and $\bar{F}^{c}_{x}$, with the fractions
of $|U_{e1}|^{2}$ and $1-|U_{e1}|^{2}$, respectively.  
This results in the appearance of the characteristic peak at $t_{pb}\sim 0.3$s, but not as
prominently as in cases (b) and (c).
\item At the early stage, (b) and (d) are indistinguishable since $\bar{P}_{\nu} \simeq 1$ for case (d),
while the time-varying probability
for case (b) also remains at $\bar{P}_{\nu} \simeq 1$ before it drops to 
$\sim 1/2$ (when the complete flavor mixture occurs \cite{Chakraborty:2011nf}) near $t_{pb} \sim 0.3$s.

\end{itemize}


\begin{figure}[ttt]
\caption{An estimation of $N(t)$ (with arbitrary normalization scale) at the Ar detector 
during the accretion phase for a $10.8 M_{\odot}$. We have used $ 9\pi/10$ as the zenith
angle at the detector. Note that the outcomes for the normal hierarchy are represented by the solid lines,
while that for the inverted hierarchy are represented by the dashed lines.} 
\centerline{\epsfig{file=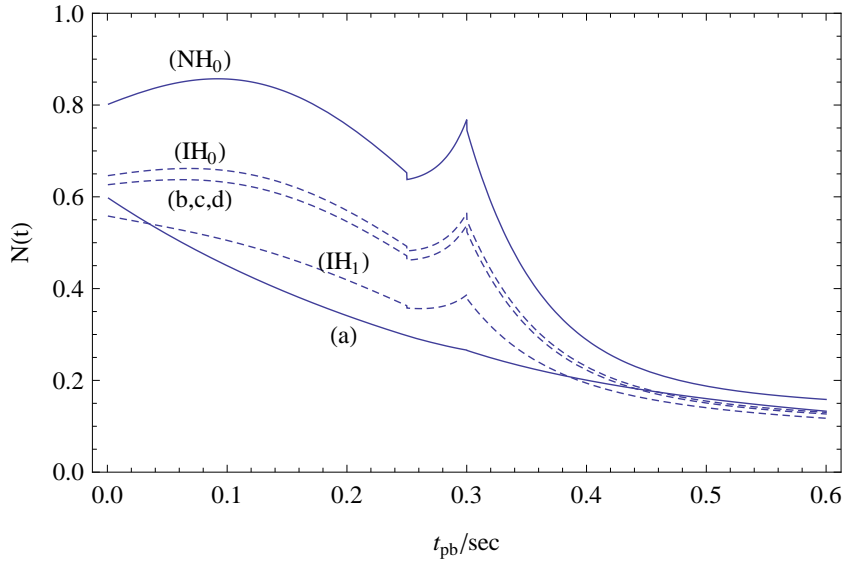,width=11 cm}}
\end{figure}

\subsection{$\nu_{e}$ events at  liquid Ar detectors}

The cross section for the charged-current reaction, $\nu_{e}+^{40}Ar \rightarrow ^{40}K^{*}+e^{-}$,
is given by \cite{Cocco:2004ac} $\sigma(\nu_{e}Ar) \simeq 3.38 \times 10^{-42}(E_{\nu}/\mbox{MeV}) \mbox{cm}^{2}$.  
We show in Fig.4 the expected time-varying behavior for the dominant $\nu_{e}$-induced events at the Ar detector. 
It is clearly seen that the case under NH with $P_{\nu} \simeq 1$ (case (a)) leads to monotonically decreasing rates, 
while a significant bump near $t_{pb}\sim 0.3s$ appears for IH with $P_{\nu} \simeq P_{s}$ 
for cases (b), (c), and (d).  
These totally distinct time structures lift the degeneracy of the $\bar{\nu}_{e}$ result between (a) and (c)
for case (iii) in the previous subsection,
and can act as a supplement to the observation of the $\bar{\nu}_{e}$ flux.
Note that one may also reason the properties of the curves in Fig.4 
by examining the details of Eq.~(\ref{eq:fen}).
In addition, three hypothetical scenarios are shown in Fig.4 as a reference, with
NH$_{0} \equiv$ (NH,$P_{\nu}=0$), IH$_{0} \equiv$ (IH,$P_{\nu}=0$), and IH$_{1} \equiv$ (IH,$P_{\nu}=1$).
So far, these three scenarios are not favored by current models in the literature.  

\subsection{Analysis with varying incident angles}

The angle of incidence at the detector is arbitrary for the SN $\nu$ fluxes,
although preferred detector locations were predicted-see, e.g., Ref.\cite{Mirizzi:2006xx}.  
A zenith angle $9\pi/10$ (mantle +core) has been adopted in our calculation.
To investigate the possible deviation from our analysis, 
we also show the results for $3\pi/5$ (mantle) and $0.01\pi$ ($\sim$ no Earth matter) in Fig.5.  
With the varying depth of the path into the Earth, the Earth effect will in principle,
modify the fluxes accordingly, as indicated by the wiggles.  The observability
of the Earth effect has been discussed elsewhere, see, e.g., Ref.\cite{Borriello:2012zc,Choubey:2010up}.  
Fig.5 suggests that the angle of incidence for the $\nu$ fluxes may modify
the details of the observed fluxes, but not the qualitative trend of the spectra.
In our analysis, we study the general time evolution of the expected fluxes,
regardless of whether or not the detailed Earth effect can be observed.

\begin{figure}[ttt]
\caption{The expected $\bar{\nu}_{e}$ flux at the WC or SC detectors for case (a), and the $\nu_{e}$ flux
arriving at the Ar detector for case (b).  Three different zenith angles 
at the detectors are adopted: $0.01\pi$ (solid),
$3\pi/5$ (dotted), and $9\pi/10$ (dashed). Despite the slight wiggles due to the Earth effect,
the general trend of the curves is unaltered. Note that the curves are plotted with arbitrary scales.} 
\centerline{\epsfig{file=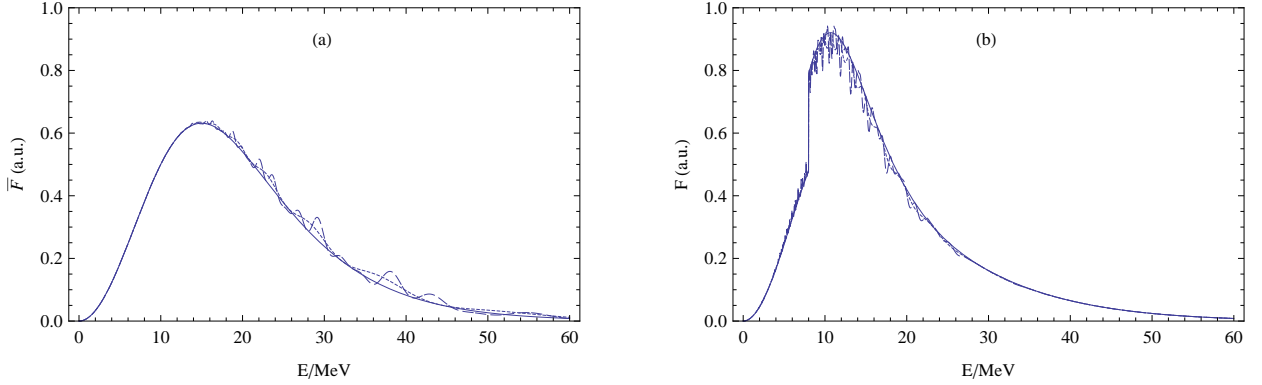,width=17 cm}}
\end{figure}

\begin{figure}[ttt]
\caption{The curves representing Eqs.(21-23) for a $18 M_{\odot}$ and a $15 M_{\odot}$ models are 
compared with the data points taken from the referential curves.} 
\centerline{\epsfig{file=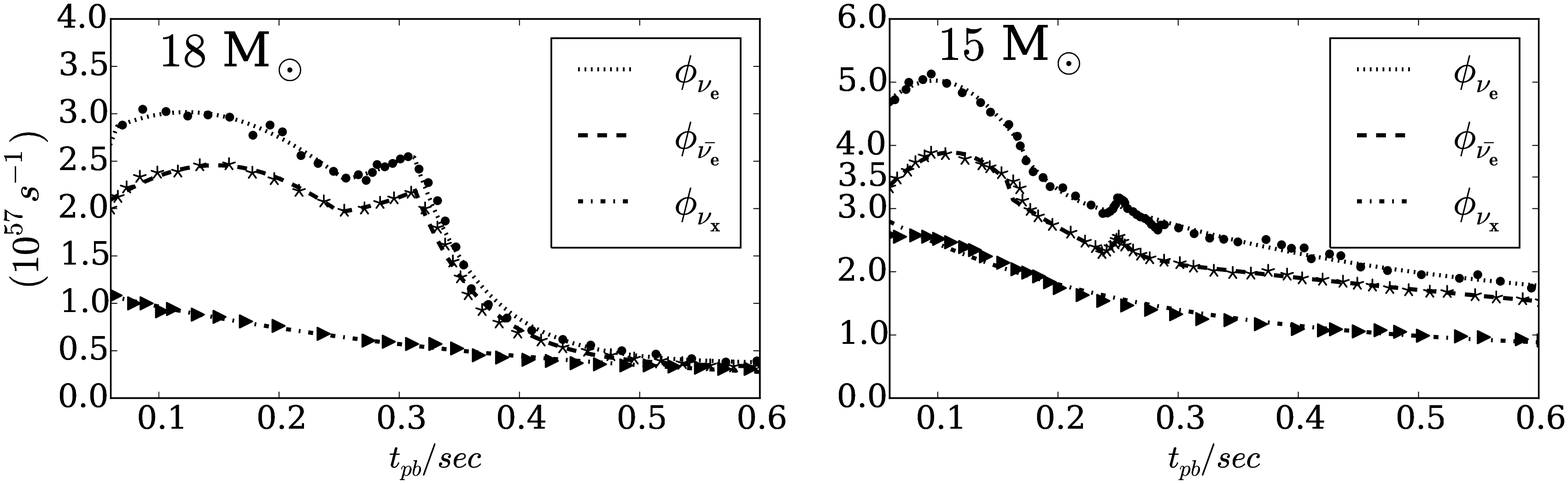,width=20 cm}}
\end{figure}

\subsection{Analysis with varying models}

In the previous analysis, we have adopted a specific, $10.8 M_{\odot}$
Fe-core progenitor model.
The time structure of the $\nu$ flux, which is model dependent, has been the main focus
of the study.
To examine the general futures and the trend of the $\nu$ luminosity obtained from other models, 
we further fit the neutrino number flux for $18 M_{\odot}$ \cite{Fischer:2009af} 
and $15 M_{\odot}$ models \cite{Borriello:2012zc} \cite{Fischer:2009af}.            
For the $18 M_{\odot}$ model, we have
\begin{equation}
\phi_{\nu_{e}} = \left\{ \begin{array}{ll}
2.52 \exp[\frac{-(t-0.13)^{2}}{0.065}], & (0 < t \leq 0.25s)  \\
               1.86+\exp[42.95t-13.30], &  (0.25s < t \leq 0.30s) \\ 
                 0.35+\exp[-16.65t+5.73], &  (0.30s <t \leq 0.60s)
                  \end{array}
                  \right.
\end{equation}
\begin{equation}
\phi_{\bar{\nu}_{e}}=\left\{\begin{array}{ll}
 1.99 \exp[\frac{-(t-0.15)^{2}}{0.051}], &  (0< t \leq 0.25s) \nonumber \\
               1.54+\exp[35.74t-11.42], &  (0.25s < t \leq 0.30s) \nonumber \\ 
                0.35+\exp[-16.24t+5.45], &  (0.30s < t \leq 0.60s)
                \end{array}
                \right.
\end{equation}
\begin{eqnarray}
\phi_{\nu_{x}}=\phi_{\bar{\nu}_{x}} = 0.13+\exp[-3.46t+0.037], & (0 < t \leq 0.60s),
\end{eqnarray}
and the $15 M_{\odot}$ model leads to
\begin{equation}
\phi_{\nu_{e}} = \left\{ \begin{array}{ll}
5.025 \exp[\frac{-(t-0.098)^{2}}{0.023}], & (0.06s < t \leq 0.16s) \nonumber \\
               2.78+\exp[-25.19t+4.38], &  (0.16s < t \leq 0.227s) \nonumber\\ 
                 2.91+\exp[39.58t-10.87], &  (0.227s <t \leq 0.253s) \nonumber\\
                 1.66+\exp[-3.76t+1.47], & (0.253s <t \leq 0.6s)  
                  \end{array}
                  \right.
\end{equation}
\begin{equation}
\phi_{\bar{\nu}_{e}}=\left\{\begin{array}{ll}
 3.89 \exp[\frac{-(t-0.098)^{2}}{0.018}], &  (0.06s< t \leq 0.177s) \nonumber \\
               0.079+\exp[-4.10t+1.79], &  (0.177s < t \leq 0.24s) \nonumber \\ 
                2.22+\exp[106.76t-27.85], &  (0.24s < t \leq 0.25s) \nonumber \\
                2.12+\exp[-60.33t+14.26], & (0.25s<t \leq 0.30s) \nonumber \\
                -0.78+\exp[-0.74t+1.29], & (0.30s <t \leq 0.6s)
                \end{array}
                \right.
\end{equation}
\begin{eqnarray}
\phi_{\nu_{x}}=\phi_{\bar{\nu}_{x}} =  0.71+\exp[-4.62t+1.01], & (0.06s <t \leq 0.6s).
\end{eqnarray}
In Fig.6, we show the fitted curves and the data points taken from the referential curves for both
the $18 M_{\odot}$ and $15 M_{\odot}$ models.

\begin{figure}[ttt]
\caption{The resultant $\bar{N}(t)$ from a $18 M_{\odot}$ and a $15 M_{\odot}$ models.} 
\centerline{\epsfig{file=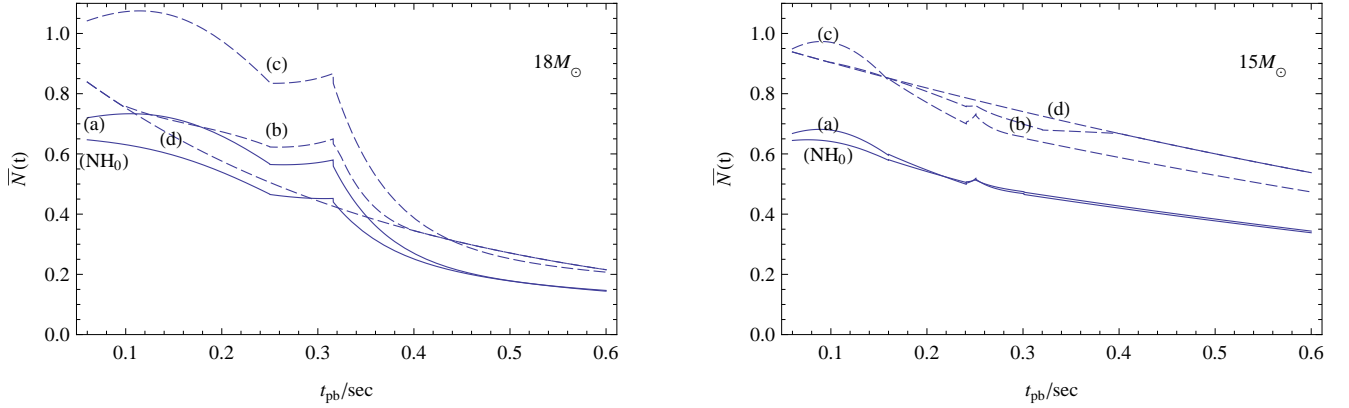,width=18 cm}}
\end{figure}

\begin{figure}[ttt]
\caption{The resultant $N(t)$ from a $18 M_{\odot}$ and a $15 M_{\odot}$ models.} 
\centerline{\epsfig{file=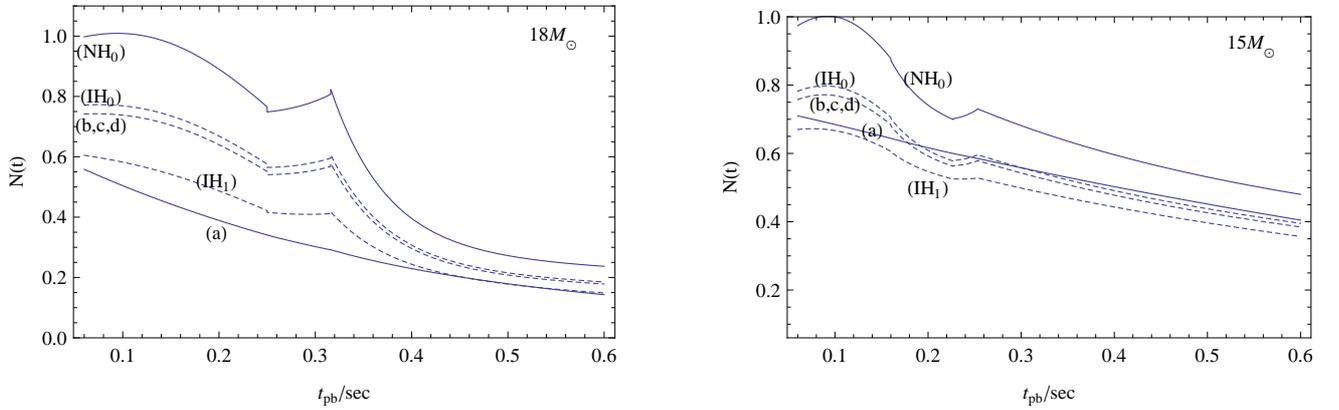,width=18 cm}}
\end{figure}

The resultant $\bar{N}(t)$ and $N(t)$ from Eq.(20) are shown in Figs.7 and 8, respectively.
One may compare Fig.3 with Fig.7 for $\bar{N}(t)$, and Fig.4 with Fig.8 for $N(t)$.
Note that for the $15M_{\odot}$ model, the peaks of both $\bar{N}(t)$ and $N(t)$ occur 
near $t_{pb} \sim 0.25$s and are less prominent as compared to other models.
It can be seen that the expected neutrino events derived from models of different SN masses
do share certain qualitative features that may be applied to the study of neutrino properties,
as outlined in the previous subsections.  However, one should also keep in mind that for complex 
physical processes such as SN events, a large portion of our knowledge
from various models in the literature still remains uncertain.  
The characteristic distinctions among models make the analysis of SN neutrinos
challenging, and more detailed models and analysis are definitely in high demand.
In fact, we should point out that
along the line of our analysis, it is difficult, if not impossible, to
apply our arguments to the results of, e.g., the $8.8 M_{\odot}$ model \cite{Fischer:2009af},
as shown in Fig.9.  
Nevertheless, a qualitative analysis based on certain types of models
such as the one discussed in this paper 
might be considered as one of the preliminary and alternative 
approaches that could pave the way for future study.

\begin{figure}[ttt]
\caption{The resultant $\bar{N}(t)$ and $N(t)$ from a $8.8 M_{\odot}$ model.} 
\centerline{\epsfig{file=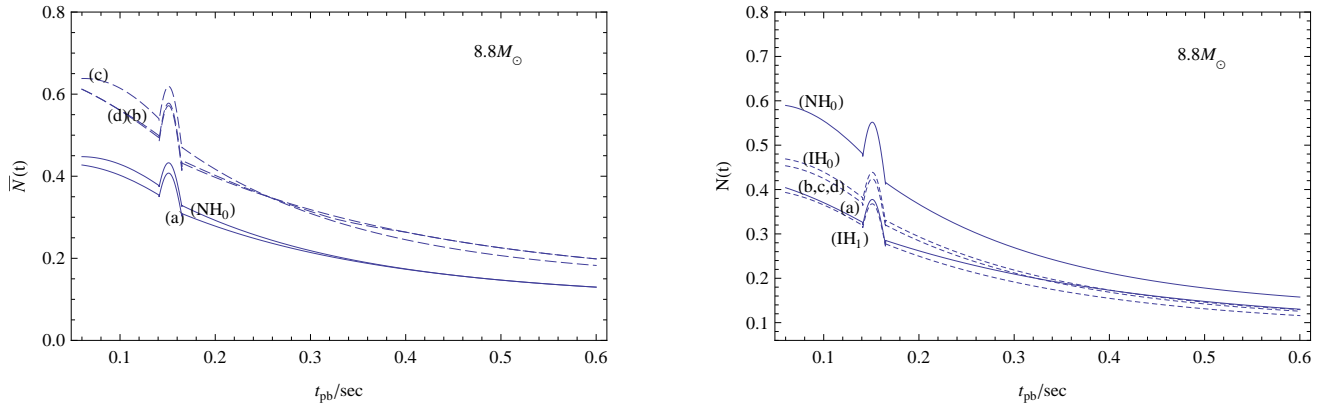,width=18 cm}}
\end{figure}

\section{Conclusion} 

The multi-angle analysis of SN neutrinos in the collective oscillations suggests
that the dense matter during the early, accretion phase may suppress the self-induced flavor conversion
and lead to time-dependent transition probabilities.  
Investigation of the neutrino signals that are modified by the influence of various collective
effects, in addition to the usual MSW and the Earth matter effects, may shed light on the 
undetermined neutrino mass hierarchy.
On the other hand, the time structures of the collective flavor conversion resulting from the
model uncertainties and the neutrino mass hierarchies during the accretion phase
may also leave signatures on the observed $\nu_{e}$ and $\bar{\nu}_{e}$ fluxes.


Intense effort has been devoted to the general study of SN neutrino flavor conversion. Related topics, such as
probing the neutrino parameters or analyzing the detectability of varied effects on the SN neutrinos, have also been
widely discussed.  To better clarify our motivation in this work and to distinguish our work from others,
we briefly outline here the aims of similar studies in the literature:  
\begin{itemize}

\item In Choubey {\it et al.} \cite{Choubey:2010up}, the expected neutrino signals resulting from varied flux models
were calculated and the signatures, especially the patterns of spectral split,
of the collective effects are examined.  
The flavor conversion during the accretion phase was also discussed.
However, the possibility that very dense ordinary matter may suppress the collective conversion during the accretion phase,
a consequence of the multi-angle analysis, was not included. 
The related discussion thus corresponds to distinguishing case (a) for the NH and case (c) for the IH
in our paper.  

\item The paper by Borriello {\it et al.} \cite{Borriello:2012zc} aimed at the observability 
of the Earth matter effect for SN neutrinos. 
The time-integrated spectra during part of the accretion phase and the cooling phase
were taken as benchmarks for the calculation.  For flavor conversion,
the complete matter suppression of the collective effect was considered,
and the neutrinos underwent only the dominant MSW conversion in the SN.
The related discussions correspond to separating case (a) and case (d) in our analysis.

\item In Serpico {\it et al.} \cite{Serpico:2011ir}, the analysis was focused on
probing the neutrino mass hierarchy with the neutrino signals at the IceCube Cherenkov detector. 
Complete matter suppression of the collective effect during the early era of the accretion phase
($t_{pb}<0.2s$) was considered. In this analysis the MSW conversion also plays the dominant role. 
The related discussion thus corresponds to distinguishing case (a) and case (d) as 
in part of our analysis ($t_{pb}<0.2s$), but based on the outcome of a different detector.

\end{itemize}

Our approach differs from the above analyses mainly in that,
as far as the influence of the collective effect is considered, 
we focus on the time-varying collective flavor 
conversion probability and the resultant
time-dependent neutrino fluxes during the accretion phase, up to $t_{pb} \sim 0.6s$.
Thus, the possible consequences due to (i) a complete suppression of the
collective effect, (ii) a partial suppression of the collective effect, and (iii) a large collective effect
are all discussed.
Given the current situation that uncertainty still remains in the mechanism of the collective effect, 
our discussion covers the scope that allows model uncertainties in the time structures
of the flavor conversion and the fluxes.  
Furthermore, our analysis was established based on a conservative assumption that 
the Earth matter effect
is beyond detectable. More precisely, observing a specific effect, such as the collective effect or the Earth matter effect,
is not the main issue in our paper and is irrelevant to our aim, although it could be crucial to other analyses.
In fact, it was suggested \cite{Borriello:2012zc} that observing the Earth effect 
under a general situation may be more
challenging than expected, and the optimistic view toward an identification of
the neutrino mass hierarchy may need to be re-evaluated.  
In a sense, this may also suggest that one should try to avoid as many as possible the factors 
that may complicate the interpretation of the signals
in probing the neutrino mass hierarchy.
Regardless of the observability of the Earth matter effect, our analysis seems in agreement with this logic
by establishing
the qualitative properties that are unaffected by whether or not the Earth matter effect can be singled out.


It should be pointed out that
the details of the time structure for the primary $\nu$ fluxes and their relative magnitude are, 
of course, model dependent. However, the general qualitative features do not vary much among the analyzed models,
as discussed in Sec. IV.  As far as our purposes are concerned,
the qualitative study should be able to provide moderate
hints as an addition to those of the numerical studies that may suffer from model uncertainties 
and the experimental details involved.  
In addition, out of the three independent effects included in our calculation, the time-dependent
collective effect plays a key role in our analysis because the variation of the expected outcome
is determined almost entirely by the not-so-well understood collective effect.  
In this regard, we further analyze the time structure of the consequences 
resulting from different models.  The results may also help shed light on 
identifying the working scenario of the collective effect in supernovae.

We have assumed equipartitioned luminosity for the neutrino flavors, as mentioned in Section II.
Note that the relative magnitude of the luminosity may not remain the same during the later cooling phase.
Even in the case of a slight deviation from the equipartition during the accretion phase, 
our qualitative analysis, which focuses on the trend of event rate in time, 
remains valid since the numerical details involved in the calculation do not alter the general pictures
of the physical observables.

Our results suggest that not only the mass hierarchy, but also the SN mechanism related to
the collective effect may, in principle, be resolved to certain extent by the observation of both $\nu$ and
$\bar{\nu}_{e}$ fluxes.  However,
this physics potential of observing SN neutrinos relies heavily on the resolution power
of a detector and proper numerical analysis.
A more detailed numerical study, which takes into consideration the possible 
consequences due to different models, variation of the luminosity strength,
and the experimental details, would certainly be in high demand and shall be discussed elsewhere.
In this work, we only estimate the qualitative properties of the observables.
Nevertheless, it is still hoped that analysis along this line would provide an alternative approach 
toward a better understanding of SN neutrinos, especially under
the current situation that the model uncertainties exist
and a comprehensive knowledge of SN physics is still lacking.

\acknowledgments 

This work is supported by the Ministry of Science and Technology of Taiwan, 
grant numbers MOST 103-2112-M-182-002 (SHC and CCH), NSC 101-2511-S-182-007 (CCH),
NSC 100-2112-M-182-001-MY3 and NSC 102-2112-M-182-001 (KCL).
We also thank Chih-Ching Chen and Tsung-Che Liu for discussions.

\appendix

\section{Derivation of $\nu$ fluxes at the detector}

We briefly summarize the derivations of Eqs.(12-15) here.  Note, as mentioned in the text, that we
have imposed the approximate conditions $P_{3e}-|U_{e3}|^{2} \leq 10^{-3}$ and 
$P_{H}\simeq \bar{P}_{H} \simeq P_{L} \simeq \bar{P}_{L} \simeq 0$ in writing
Eqs.(12-15) from part of the following derivations.

The original fluxes $F_{e}^{0}$ and $F_{x}^{0}$ in Eq.(69) of Ref.\cite{Dighe:1999bi},
\begin{equation}
F^{D}_{e} \simeq F_{e}+P_{H}(P_{2e}-|U_{e2}|^{2})(1-2P_{L})(F_{e}^{0}-F_{x}^{0}),
\end{equation}
should be respectively replaced by our $F_{e}^{c}$ (Eq.(2)) and $F_{x}^{c}$ (Eq.(3))
to account for the collective effect:
\begin{equation}\label{eq:A2}
F^{D}_{e} \simeq F_{e}+P_{H}(P_{2e}-|U_{e2}|^{2})(1-2P_{L})(F_{e}^{c}-F_{x}^{c}),
\end{equation}
where $F_{e}$, as given by Eq.(6), is the $\nu_{e}$ flux arriving at Earth.
Eq.~(\ref{eq:A2}) leads directly to $F_{e}^{D}$ under the normal hierarchy, as given by
Eq.(12), if one imposes the approximate conditions.

In addition, since the conversion of $\nu_{e}$ is independent of $P_{H}$ under the inverted hierarchy,
we may simply set $P_{H}=1$ in Eq.(71) of Ref.\cite{Dighe:1999bi} and use their Eq.(69) to obtain the expression
for $F_{e}^{D}$ under the inverted hierarchy.  This leads to Eq.(13) if one again replaces
$F_{e}^{0}-F_{x}^{0}$ by $F_{e}^{c}-F_{x}^{c}$.

As for the $\bar{\nu}_{e}$ flux at the detector under the normal hierarchy,
we use Eq.(79) of Ref.\cite{Dighe:1999bi} and replace $F_{\bar{e}}^{0}$ 
and $F_{x}^{0} (=F_{\bar{x}}^{0})$ respectively
by $\bar{F}_{e}^{c}$ (Eq.(4)) and $\bar{F}_{x}^{c}$ (Eq.(5)). 
This leads to the expression of $\bar{F}_{e}^{D}$ in Eq.(14).
Note that the original flux of $\nu_{x}$ is usually set to
be identical to that of $\bar{\nu}_{x}$, $F_{x}^{0}=F_{\bar{x}}^{0}$.

Finally, for the inverted hierarchy, the $\bar{\nu}_{e}$ spectrum is the same as the normal one
except for a suppression factor $\bar{P}_{H}$.  In this case, $\bar{F}_{e}^{D}$ in Eq.(15) can be
derived from Eq.(106) of Ref.\cite{Dighe:1999bi} by setting $D_{2}$ at 
the location where the $\nu$ flux enters the Earth
so that $F_{\bar{e}}^{D2}=\bar{F}_{e}$ and $\bar{P}_{1e}^{(2)}=|U_{e1}|^{2}$.
Explicitly, one may refer to, e.g., Eq.(27) of Ref. \cite{Lunardini:2001pb}.
Following this line, one reaches Eq.(15) if $F_{\bar{e}}^{0}-F_{x}^{0}$ 
there in Ref.\cite{Dighe:1999bi} or \cite{Lunardini:2001pb}
is replaced by $\bar{F}_{e}^{c}-\bar{F}_{x}^{c}$.


\begin{thebibliography}{}

  


\bibitem{Pantaleone:1992eq} 
  J.~T.~Pantaleone,
  Phys.\ Lett.\ B {\bf 287}, 128 (1992)
  Y.~Z.~Qian and G.~M.~Fuller,
  Phys.\ Rev.\ D {\bf 51}, 1479 (1995)
  H.~Duan, G.~M.~Fuller and Y.~-Z.~Qian,
  Phys.\ Rev.\ D {\bf 74}, 123004 (2006)
  H.~Duan, G.~M.~Fuller, J. Carlson and Y.~-Z.~Qian,
  Phys.\ Rev.\ D {\bf 74}, 105014 (2006) 
  S.~Hannestad, G.~G.~Raffelt, G.~Sigl and Y.~Y.~Y.~Wong,
  Phys.\ Rev.\ D {\bf 74}, 105010 (2006) 

  
  
  
    \bibitem{Duan:2010bg} 
  H.~Duan, G.~M.~Fuller and Y.~-Z.~Qian,
  Ann.\ Rev.\ Nucl.\ Part.\ Sci.\  {\bf 60}, 569 (2010)
  

  
  \bibitem{Wolfenstein:1977ue+Mikheev:1986gs} 
  L.~Wolfenstein,
  Phys.\ Rev.\ D {\bf 17}, 2369 (1978)
   S.~P.~Mikheev and A.~Y.~.Smirnov,
  Sov.\ J.\ Nucl.\ Phys.\  {\bf 42}, 913 (1985)
  [Yad.\ Fiz.\  {\bf 42}, 1441 (1985)]
  
  
\bibitem{Fogli:2007bk} 
  G.~L.~Fogli, E.~Lisi, A.~Marrone and A.~Mirizzi,
  JCAP {\bf 0712}, 010 (2007)



\bibitem{Dasgupta:2009mg} 
  B.~Dasgupta, A.~Dighe, G.~G.~Raffelt and A.~Y.~.Smirnov,
  Phys.\ Rev.\ Lett.\  {\bf 103}, 051105 (2009)
  
 \bibitem{Raffelt:2010zza} 
  G.~G.~Raffelt,
  Prog.\ Part.\ Nucl.\ Phys.\  {\bf 64}, 393 (2010). 
  
  
\bibitem{An:2012eh} 
  F.~P.~An {\it et al.}  [DAYA-BAY Collaboration],
  Phys.\ Rev.\ Lett.\  {\bf 108}, 171803 (2012)  
  
  \bibitem{Ahn:2012nd} 
  J.~K.~Ahn {\it et al.}  [RENO Collaboration],
  Phys.\ Rev.\ Lett.\  {\bf 108}, 191802 (2012)
  

 \bibitem{EstebanPretel:2008ni} 
  A.~Esteban-Pretel, A.~Mirizzi, S.~Pastor, R.~Tomas, G.~G.~Raffelt, P.~D.~Serpico and G.~Sigl,
  Phys.\ Rev.\ D {\bf 78}, 085012 (2008) 
  
  
  \bibitem{Chakraborty:2011gd} 
  S.~Chakraborty, T.~Fischer, A.~Mirizzi, N.~Saviano and R.~Tomas,
  Phys.\ Rev.\ D {\bf 84}, 025002 (2011)
  
  
\bibitem{Chakraborty:2011nf} 
  S.~Chakraborty, T.~Fischer, A.~Mirizzi, N.~Saviano and R.~Tomas,
  Phys.\ Rev.\ Lett.\  {\bf 107}, 151101 (2011)
  
\bibitem{Sarikas:2011am} 
  S.~Sarikas, G.~G.~Raffelt, L.~Hudepohl and H.~-T.~Janka,
  Phys.\ Rev.\ Lett.\  {\bf 108}, 061101 (2012)  
  
  \bibitem{data}
  J. Beringer et al. (Particle Data Group), Phys. Rev. D 86, 010001 (2012)

  \bibitem{Totani:1997vj} 
  T.~Totani, K.~Sato, H.~E.~Dalhed and J.~R.~Wilson,
  Astrophys.\ J.\  {\bf 496}, 216 (1998)
 
  \bibitem{Keil:2002in} 
  M.~T.~.Keil, G.~G.~Raffelt and H.~-T.~Janka,
  Astrophys.\ J.\  {\bf 590}, 971 (2003)
  
   
  \bibitem{Serpico:2011ir} 
  P.~D.~Serpico, S.~Chakraborty, T.~Fischer, L.~Hudepohl, H.~-T.~Janka and A.~Mirizzi,
  Phys.\ Rev.\ D {\bf 85}, 085031 (2012)
  
  
  \bibitem{Fogli:2008pt} 
  G.~L.~Fogli, E.~Lisi, A.~Marrone, A.~Mirizzi and I.~Tamborra,
  Phys.\ Rev.\ D {\bf 78}, 097301 (2008)
  
  \bibitem{Dighe:1999bi} 
  A.~S.~Dighe and A.~Y.~Smirnov,
  Phys.\ Rev.\ D {\bf 62}, 033007 (2000)
  
  
  \bibitem{Chiu:2008sa} 
  S.~-H.~Chiu and T.~-K.~Kuo,
  Phys.\ Rev.\ D {\bf 73}, 033007 (2006)
   S.~-H.~Chiu,
  Phys.\ Rev.\ D {\bf 76}, 045004 (2007)
  S.~-H.~Chiu,
  Mod.\ Phys.\ Lett.\ A {\bf 24}, 2741 (2009)
  
  
  
  \bibitem{deHolanda:2004fd} 
  P.~C.~de Holanda, W.~Liao and A.~Y.~Smirnov,
  Nucl.\ Phys.\ B {\bf 702}, 307 (2004)
  
  \bibitem{Freund:1999vc} 
  M.~Freund and T.~Ohlsson,
  Mod.\ Phys.\ Lett.\ A {\bf 15}, 867 (2000)
  
  \bibitem{Cadonati:2000kq} 
  L.~Cadonati, F.~P.~Calaprice and M.~C.~Chen,
  Astropart.\ Phys.\  {\bf 16}, 361 (2002)
  
  
  \bibitem{Cocco:2004ac} 
  A.~G.~Cocco, A.~Ereditato, G.~Fiorillo, G.~Mangano and V.~Pettorino,
  JCAP {\bf 0412}, 002 (2004)
  
  \bibitem{Mirizzi:2006xx} 
  A.~Mirizzi, G.~G.~Raffelt and P.~D.~Serpico,
  JCAP {\bf 0605}, 012 (2006)
  
  
  \bibitem{Borriello:2012zc} 
  E.~Borriello, S.~Chakraborty, A.~Mirizzi, P.~D.~Serpico, and I.~Tamborra,
  Phys.\ Rev.\ D {\bf 86}, 083004 (2012)
 
  \bibitem{Choubey:2010up} 
  S.~Choubey, B.~Dasgupta, A.~Dighe and A.~Mirizzi,
  arXiv:1008.0308 [hep-ph]
   
  \bibitem{Fischer:2009af} 
  T.~Fischer, S.~C.~Whitehouse, A.~Mezzacappa, F.~-K.~Thielemann and M.~Liebendorfer,
  Astron.\ Astrophys.\  {\bf 517}, A80 (2010)
 
\bibitem{Lunardini:2001pb} 
  C.~Lunardini and A.~Y.~Smirnov,
  Nucl.\ Phys.\ B {\bf 616}, 307 (2001)
  [hep-ph/0106149].
  
   
\end{thebibliography}
\end{document}